\pdfoutput=1
\documentclass[12pt]{iopart}
\usepackage[utf8]{inputenc}
\usepackage{iopams}
\usepackage{amsfonts}
\usepackage{graphicx}

\newcommand{\muG}{$\mu$Gal}

\begin{document}
\maketitle
\title{Comment on ``Measurement of the speed-of-light perturbation of free-fall absolute gravimeters''}
\author{V D Nagornyi}
\address{Metromatix, Inc., 111B Baden Pl, Staten Island, NY 10306, USA}
\ead{vn2@member.ams.org}

\begin{abstract}
The paper (Rothleitner et al. 2014 Metrologia 51, L9) reports on the measurement of the \emph{speed-of-light} perturbation in absolute gravimeters. The conclusion that the perturbation reaches only $\frac23$ of the commonly accepted value violates the fundamental limitation on the maximum speed of information transfer. The conclusion was deluded by unaccounted parasitic perturbations, some of which are obvious from the report. 
\end{abstract}
%
%
\pagestyle{empty}
%
%
%
%
\section{The ``$\frac23$ result'' and its interpretations}
\begin{figure}[h]
\centering
\includegraphics[height=50mm]{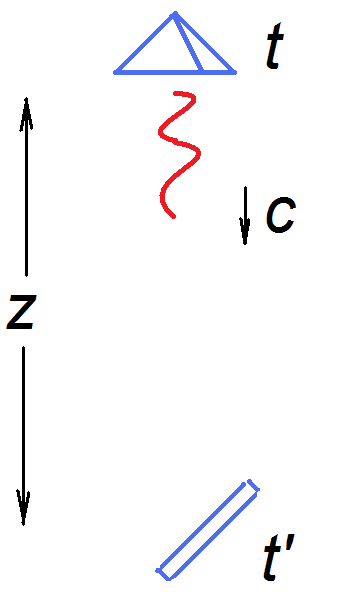}
\label{fig_delay}
\caption{The delay of the laser beam.}
\end{figure}

The \emph{speed-of-light} perturbation in absolute gravimeters is due to the fact that the laser beam reflected from the test mass delivers the information on its position with some delay. The light reflected at the moment $t$ reaches the beam splitter at a later moment $t'$, so that
\begin{equation}
\label{eq_delay}
t' = t + \frac{z}{c},
\end{equation}
where $z$ is the distance traveled by the light, $c$ is the speed of light (fig.\ref{fig_delay}). 

The paper \cite{rothleitner2014} reports on the experiment which agrees better with the formula
\begin{equation}
\label{eq_delay23}
t' = t + \frac23 \; \frac{z}{c}.
\end{equation}
As there are only two values: $z$ and $c$ involved in the correction (\ref{eq_delay}), the only two possible  interpretations of the result (\ref{eq_delay23}) would be
\begin{equation}
\label{eq_interp}
\frac23 \; \frac{z}{c} = \frac{\frac23 \; z}{c} 
\;\;\;\;\;\;\;\;\;\;\textrm{or}\;\;\;\;\;\;\;\;\;\;\;\;
\frac23 \; \frac{z}{c} = \frac{z}{\frac32 \; c}.
\end{equation}
The first interpretation means that the path traveled by the light is one-third less than the distance to the beam splitter. Another equivalent interpretation means that the information was delivered  50\% faster than the speed of light.
\section{The experiment}
The experiment compared the values of the gravity acceleration obtained at different drop lengths. Based on the fact that the \emph{speed-of-light} perturbation depends on the drop length, the analysis of the data has revealed that the observed dependence is better described by the commonly accepted formula, if it's multiplied by the factor $\frac23$. The experiment included two sets of lengths resulted from varying either the first fringe (FF) or the last fringe (LF) of the drop. The longest drop was the same in both sets spanning 30 cm covered by the test mass in 0.222 s.
The major difficulty of the experiment is that almost all instrumental perturbations also depend on the drop length, so their influences had to be excluded as part of the data analysis. Though a lot of efforts was put into identifying and eliminating other perturbances, there are evidences that the analysis is incomplete.
\section{Some problems with the data analysis}
The formula for the \emph{speed-of-light} perturbation of the acceleration \cite{robertsson2005} used in the analysis yields 14.18 \muG\footnote{1 \muG = $10^{-8}$ ms$^{-2}$} for the longest drops in both sets, so the one-third of the correction is 4.73 \muG. However, on the fig.4 of \cite{rothleitner2014} (FF set) this difference reaches only 2.2 \muG, and on the fig.5 (LF set) it is only 1.1 \muG. Besides being much lower than the right value, the LF and FF results disagree between themselves, even though the longest drops are the same in both sets. These facts indicate the presence of large non-\emph{speed-of-light} perturbations not accounted in the analysis.

\begin{figure}
\centering
\begin{tabular}{@{}cccc@{}}
0  & $\pi$/4 & $\pi$/2 & 3$\pi$/4 \\
\includegraphics[height=25mm]{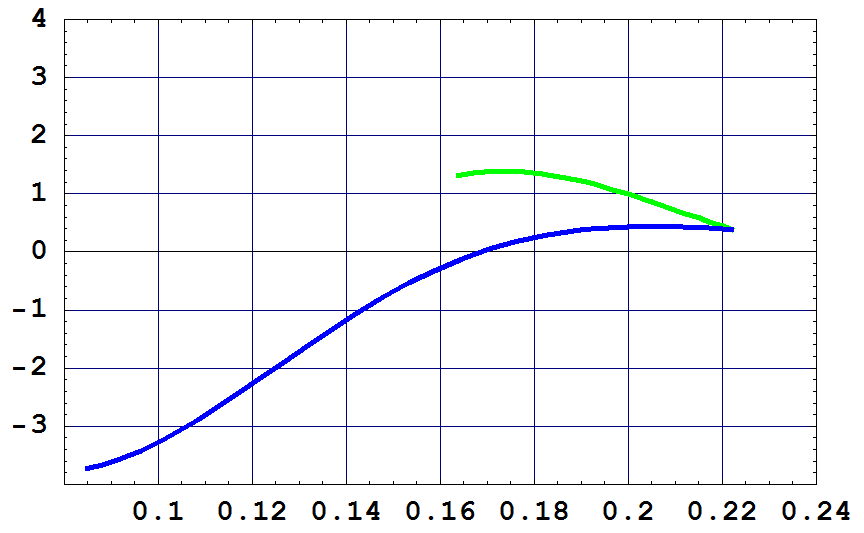} &
\includegraphics[height=25mm]{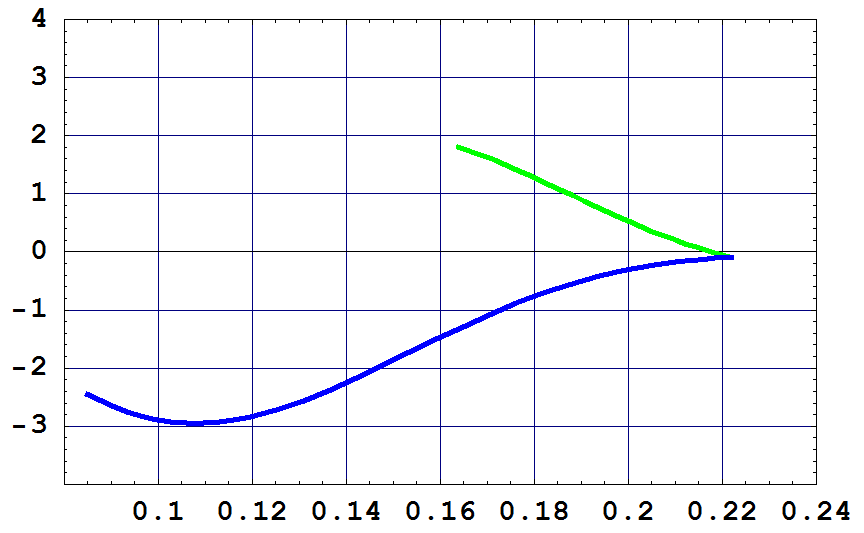} &
\includegraphics[height=25mm]{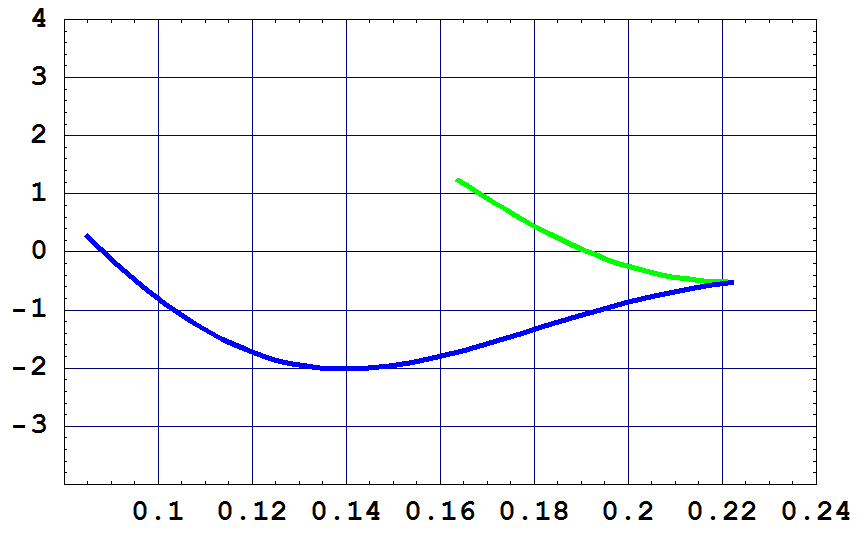} &
\includegraphics[height=25mm]{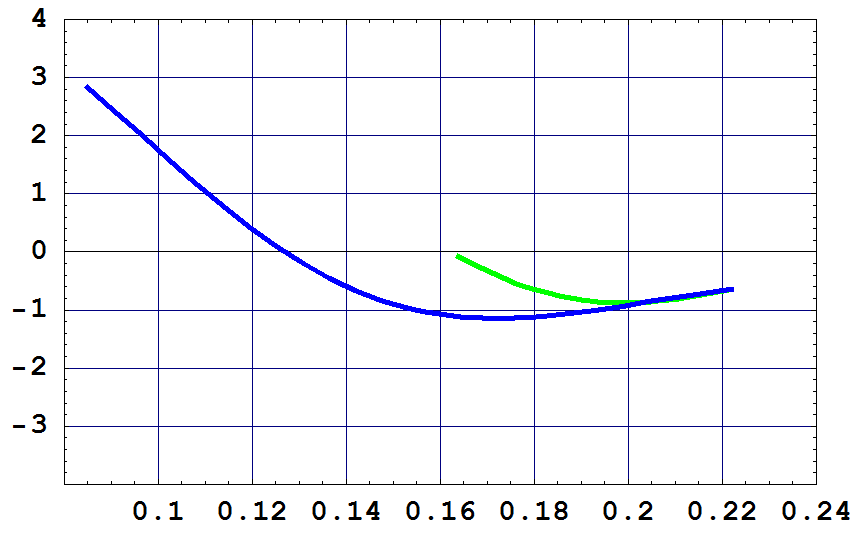} \\ 
$\pi$  & 5$\pi$/4 & 3$\pi$/2 & 7$\pi$/8 \\
\includegraphics[height=25mm]{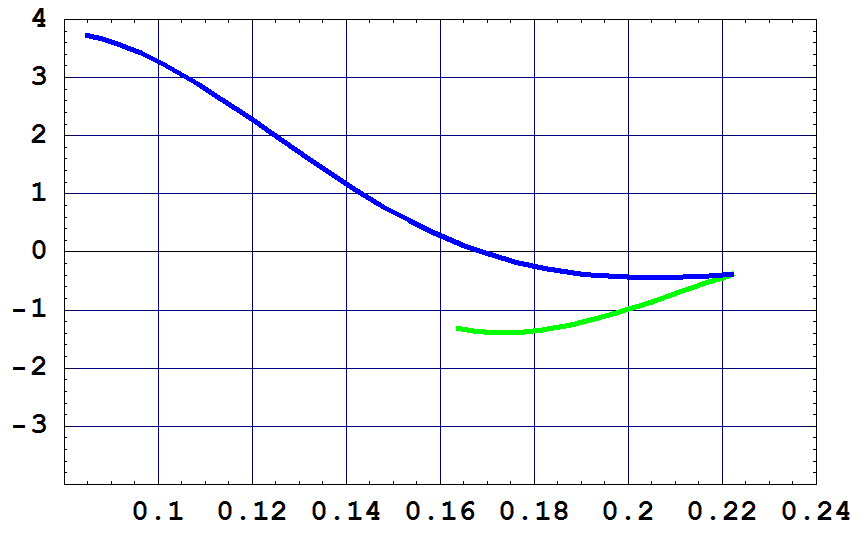} &
\includegraphics[height=25mm]{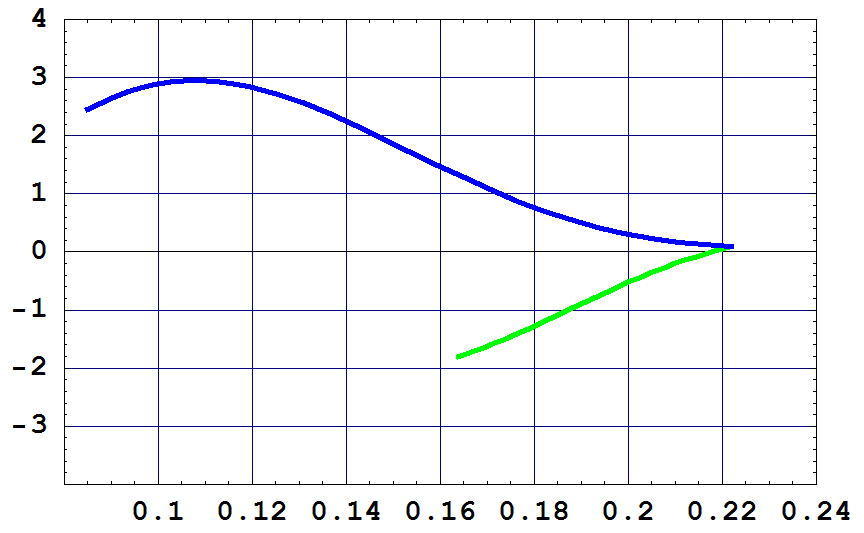} &
\includegraphics[height=25mm]{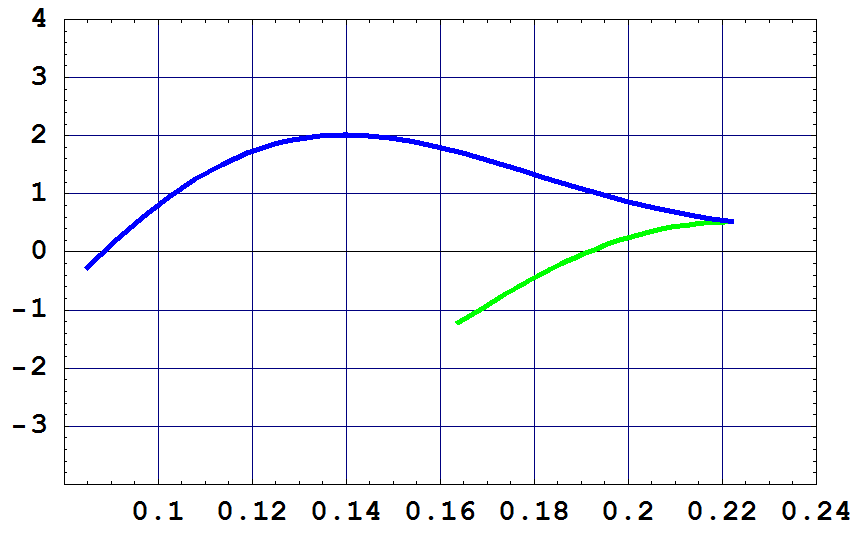} &
\includegraphics[height=25mm]{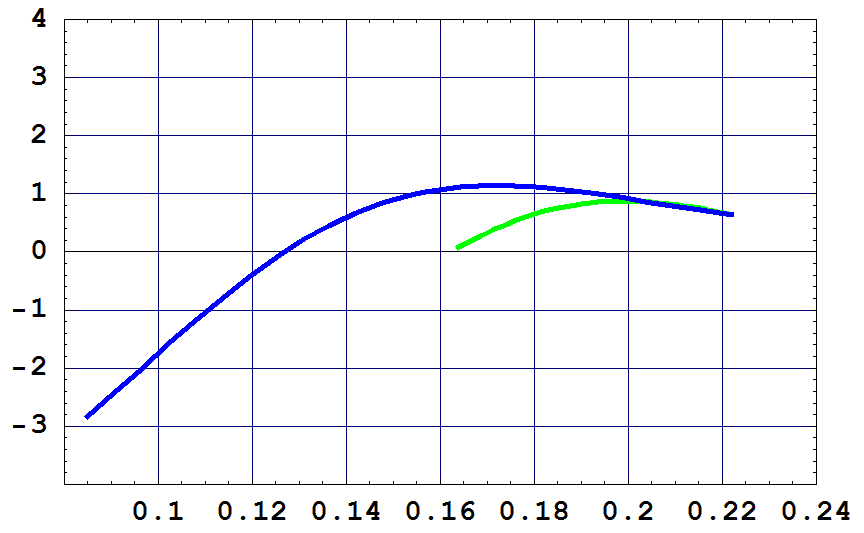} \\ 
\end{tabular}
\label{fig_fit}
    \caption{Perturbation of the acceleration (in \muG)   due to the harmonic of 0.025~nm, 7~Hz for the drop times (in seconds) of the FF (long line) and LF (short line) experiments.}
\end{figure}

According to \cite{rothleitner2014}, one of the harmonics present in the stacked residuals has the amplitude of 0.025 nm and the frequency of 7 Hz. The fig.\ref{fig_fit} shows the perturbation caused by the harmonic found by fitting to it a quadratic parabola using the same data points as in the original experiment \cite{rothleitner2014}. The calculations show that the perturbation can reach 4 \muG$\,$ at the shortest drops. This value is in good agreement with the result of \cite{svitlov2012} (section 4.2.1) and about an order of magnitude higher than 0.46 \muG$\,$ estimated in the report (probably, for the longest drop only). Exact influence of the harmonic on the experiment depends on the initial phase (fig.\ref{fig_fit}). No phase, however, leaves any chance to the ``$\frac23$ result'' to withstand.
\section{Conclusions}
The paper \cite{rothleitner2014} presents the first attempt of the  measurement of the \emph{speed-of-light} perturbation in absolute gravimeters. The result that the perturbation reaches only $\frac23$ of the commonly accepted value requires the information to be delivered faster than the speed of light. The result is caused by other perturbations not accounted in the analysis. We have analyzed only one perturbation and found that its magnitude is about 10 times higher than reported in \cite{rothleitner2014}. It's possible that the analysis of other perturbations can also benefit from the review by other researchers. Because few of them possess technical means to reproduce the experiment, it's desirable that the authors of \cite{rothleitner2014} (should they stand by their conclusion) publish a more detailed description of the data processing along with the raw data used in the analysis\footnote{Many modern journals including \emph{Metrologia} allow for online publishing of such supplementary materials.}. This would allow for independent verification of the result and enhancement of future experiments.

%
%
%

\section*{References}


\end{document}